\documentclass[12pt]{article} 
\usepackage{ol2}
\usepackage[draft]{hyperref}
\usepackage{amsmath}
\begin{document}
\title{Electrostatic theory for designing lossless negative permittivity metamaterials}
\author{Yong Zeng, Qi Wu and Douglas H. Werner}
\address{Department of Electrical Engineering, Pennsylvania
State University, State College, Pennsylvania 16802}
\begin{abstract}
In this Letter, we develop an electrostatic theory for designing
bulk composites with effective lossless negative permittivities.
The theory and associated design procedure are validated by
comparing their predictions with those of rigorous full-wave
simulations. It is demonstrated that the excitation of the
Fr\"{o}hlich mode (the first-order surface mode) of the
constitutive nanoparticles plays a key role in achieving negative
permittivities with compensated losses.
\end{abstract}

\newpage
Metamaterials with simultaneous negative permeability and
permittivity have attracted much attention in the last decade
\cite{marques,solymar,zhang,smith}. Such novel media are absent in
nature and have potential applications such as perfect imaging
\cite{john} and optical cloaking \cite{pendry,leonhardt,schurig}
as well as other transformation optical devices \cite{rahm,kwon}.
Unfortunately, the intrinsic ohmic loss of plasmonic
nanostructures seriously deteriorates their performance. To
alleviate metal absorption loss, metamaterials containing gain
media have been proposed
\cite{Bergman,noginov2,zheludev,Noginov,Oulton,fang,stockman,gordon}.
For instance, a recent experiment found that 44-nm-diameter coated
spheres with a gold core and dye-doped silica shell can amplify
light at a wavelength of 531 nm \cite{Noginov}.

In accordance with the definition of refractive index $n$, a
negative value of $n$ requires simultaneous negative dielectric
permittivity $\epsilon$ and magnetic permeability $\mu$. Searching
for materials with negative dielectric constant and compensated
loss is therefore extremely important. As the first attempt, Fu
\textit{et al.} suggested metamaterials comprised of binary metal
and semiconductor quantum dots (with optical gain)
\cite{fu,Bratkovsky,Ponizovskaya}, numerically demonstrating that
these composites indeed possess an effective negative real-valued
$\epsilon$ at specific frequencies. However, the underlying
physical mechanisms that governs this class of novel materials
have not been fully explained, nor is there an easy-to-use design
procedure. In this Letter, we employ nanopaticles to build up a
bulk composite with lossless negative permittivity, and develop an
electrostatic theory to illustrate the underlying physics. We find
that the crucial requirement is the excitation of the first-order
surface mode of the constitutive nanoparticles. Consequently,
these particles present localized surface plasmon singularities
\cite{lawandya}, which further result in extremely high
enhancement of local field intensity as well as surface-enhanced
Raman scattering \cite{li}.

We start by considering a coated sphere with inner radius $r_{1}$
and outer radius $r_{2}$, surrounded by a homogenous medium that
has a positive real-valued permittivity $\epsilon_{m}$. Its
electromagnetic response can be described by a rigorous dynamic
theory developed almost 60 years ago \cite{aden}. When the
incident wavelength is much larger than the size of the particle,
this coated sphere can be approximated as an ideal dipole with
polarizability \cite{bohren}
\begin{equation}
\alpha=4\pi
r_{2}^{3}\frac{(\epsilon_{2}-\epsilon_{m})(\epsilon_{1}+2\epsilon_{2})+\rho(\epsilon_{1}-\epsilon_{2})(\epsilon_{m}+2\epsilon_{2})}{(\epsilon_{2}+2\epsilon_{m})(\epsilon_{1}+2\epsilon_{2})+2\rho(\epsilon_{2}-\epsilon_{m})(\epsilon_{1}-\epsilon_{2})},
\label{eq1}
\end{equation}
where $\rho=r_{1}^{3}/r_{2}^{3}$ is the fraction of the total
particle volume occupied by the inner sphere material, while
$\epsilon_{1}$ and $\epsilon_{2}$ are the permittivity of the core
and shell, respectively. We further recall that a homogeneous
sphere in the electrostatics approximation can also be treated as
an ideal dipole with polarizability \cite{bohren}
\begin{equation}
\alpha=4\pi
r^{3}\frac{\epsilon-\epsilon_{m}}{\epsilon+2\epsilon_{m}},
\label{eq2}
\end{equation}
with $r$ being the radius and $\epsilon$ being the permittivity.
Consequently, a coated sphere can be approximated as an effective
homogeneous sphere with an equivalent permittivity given by
\begin{equation}
\epsilon_{c}=\epsilon_{2}\frac{\epsilon_{1}(1+2\rho)+2\epsilon_{2}(1-\rho)}{\epsilon_{1}(1-\rho)+\epsilon_{2}(2+\rho)},
\label{eq3}
\end{equation}
which bears a resemblance to the standard Maxwell-Garnett mixing
rule for a heterogeneous medium where homogeneous spheres are
dilutely mixed into an isotropic host environment \cite{jackson}.
It should be emphasized that this equivalent permittivity can be
derived directly by solving Laplace's equation of electrostatics
together with proper boundary conditions \cite{jackson}. For a
bulk material consisting of these nanoparticles, its effective
electrostatic permittivity is determined by the Clausius-Mossotti
relation, also known as the Lorentz-Lorenz formula
\cite{jackson,yannopapas,wheeler},
\begin{equation}
\epsilon_{b}=\epsilon_{m}\frac{4\pi\:r_{2}^{3}+2f\alpha}{4\pi\:r_{2}^{3}-f\alpha},
\label{eq4}
\end{equation}
where $f$ is the filling fraction of the composite. Substituting
$\alpha$ with Eq. (\ref{eq2}), the above expression can be
rewritten as
\begin{equation}
\epsilon_{b}=\epsilon_{m}\frac{\epsilon_{c}(1+2f)+2\epsilon_{m}(1-f)}{\epsilon_{c}(1-f)+\epsilon_{m}(2+f)}.
\label{eq5}
\end{equation}
Once again, this effective permittivity has an expression
identical to the standard Maxwell-Garnett formula.

A lossless dielectric constant $\epsilon_{b}$ must be entirely
real, which leads to the condition that
\begin{equation}
\mathrm{Im}(\epsilon_{c})=0 \label{eq6}
\end{equation}
implying that the absorption cross section of the nanoparticle is
always zero \cite{jackson}. The particle therefore must contain a
medium with gain to compensate for the ohmic losses of the metal.
Requiring $\epsilon_{b}$ to be negative results in
\begin{equation}
\epsilon_{m}\frac{2+f}{f-1}<\mathrm{Re}(\epsilon_{c})<\epsilon_{m}\frac{2f-2}{1+2f}.
\label{eq7}
\end{equation}
The real part of $\epsilon_{c}$ is negative since the filling
fraction $f$ is always smaller than unity. Furthermore, the
left-hand expression is a strictly decreasing function with a
maximum of $-2\epsilon_{m}$ at $f=0$, while the expression on the
right-hand side is a monotonically increasing function with a
minimum of $-2\epsilon_{m}$. $\mathrm{Re}(\epsilon_{c})$ therefore
should be tuned to around $-2\epsilon_{m}$ so that the filling
fraction $f$ can be modest. The combination of Eq. (\ref{eq6}) and
Eq. (\ref{eq7}), i.e., $\epsilon_{c}=-2\epsilon_{m}$, is actually
the Fr\"{o}hlich mode (the first-order surface mode) condition for
the individual particle \cite{fro}. More specifically, the
denominator of the electric dipolar polarizability (see Eq.
(\ref{eq2})) goes to zero at the Fr\"{o}hlich frequency. The
first-order scattering coefficient of the nanoparticle is
therefore infinite without accounting for saturation effects,
which results in an infinite scattering cross section. This
phenomenon was recently named the surface plasmon singularity
\cite{noginov2,lawandya,bohren}.

Although the above conclusion is derived from coated spheres, it
can be extended to general nanoparticles with arbitrary shapes. As
suggested by the Clausius-Mossotti relation, the electric dipolar
polarizability $\alpha$ should be large enough to achieve a
negative real-valued $\epsilon_{b}$ which is weakly dependent on
the filling fraction. We recall that $\alpha$ is infinite at the
Fr\"{o}hlich frequency. Therefore we need to excite the
first-order surface mode of the constitutive nanoparticles to tune
the effective permittivity of the bulk composite to be real and
negative. It should be mentioned that, accompanied by the surface
mode, the electromagnetic field is strongly localized around the
nanoparticle, which further greatly enhances the Raman scattering
signals \cite{li}.

We can now design bulk composites with negative real-valued
dielectric constants. A first example consists of silver coated
spherical semiconductor quantum dots (QDs). Specifically, we
employ similar QDs to those in Ref.\cite{fu} whose permittivity is
described by a realistic two-level model
\cite{fu,Bratkovsky,Ponizovskaya},
\begin{equation}
\epsilon_{\infty}-\frac{\beta}{\omega_{0}-\omega-i\gamma},
\label{eq8}
\end{equation}
where the parameters are given by $\epsilon_{\infty}=12.8$,
$\hbar\omega_{0}=1.5$ eV ($\lambda=827$ nm), $\hbar\gamma=1$ meV
and $\beta/\gamma=100$ \cite{fu}. In addition, the relative
dielectric constant of silver is fitted by the Drude model of
$4.56-\omega_{p}^{2}/(\omega^{2}+i\gamma\omega)$, with
$\omega_{p}=1.4\times10^{16} s^{-1}$ and $\gamma=1.0\times10^{14}
s^{-1}$ \cite{palik}. Furthermore, the radius of the spherical QD
is 5 nm, and the silver shell has a thickness of 2.25 nm. Using
Eq. (\ref{eq3}), the permittivity of an individual particle
$\epsilon_{c}$ is computed and the result is plotted in Fig. 1(a).
Around a wavelength of 822 nm, the curve of
$\mathrm{Im}(\epsilon_{c})$ passes through zero (marked with the
dotted line), and $\mathrm{Re}(\epsilon_{c})$ is close to $-2.0$,
showing that the first-order surface mode is excited. To elucidate
its characteristics, the corresponding local optical field is
plotted in Fig. 1(c). It shows a uniform field distribution inside
the core and hot spots with strong electric field at the poles
outside the shell. In other words, light is significantly enhanced
around the nanoshell. Using Eq. (\ref{eq5}), we further compute
the effective permittivity $\epsilon_{b}$ of a bulk composite with
a filling fraction of $0.1$; Fig. 1(b) shows the result. Indeed, a
lossless $\epsilon_{b}=-2.0$ is found at 821.8 nm wavelength. To
verify the electrostatic result, we consider the coated spheres in
a simple cubic lattice with lattice spacing of 20 nm. A full-wave
finite-element simulation is then used to calculate the
corresponding scattering matrix \cite{comsol}, and the
transmission/reflection method is finally applied to extract the
effective permittivity \cite{smith2,simocski}. The result is
plotted in Fig. 1(b) with dotted curves, and is found to be in
perfect agreement with its analytical counterpart.

A second example is inspired by the active gold nanobox studied in
Ref.\cite{li}. By embedding a gain medium in the core, this cubic
particle is found to extremely enhance local electric intensity so
that single-molecule detection is achievable through
surface-enhanced Raman scattering \cite{li}. Although the
underlying mechanism was not demonstrated in the original
literature, it is quite possible that it is related to the
first-order surface mode of the active nanobox. To prove our
theory, we consider a similar gold nanoshell with almost identical
parameters to the nanobox. More specifically, the gain medium has
a refractive index of $1.33-0.1437i$, the refractive index of the
surrounding medium is 1.33 (therefore $\epsilon_{m}=1.77$), and
the permittivity of the gold is described by a Drude-Lorentz model
\cite{vial}. In addition, the core radius is 16 nm and the gold
shell thickness is 2 nm. The nonlocal effect of gold is not
considered here since it does not influence our results
qualitatively \cite{mcMahon}. Similar to the first design, we
compute the effective permittivities analytically and numerically,
and the results are summarized in Fig. 2. Around a wavelength of
789 nm, the complex $\epsilon_{c}$ is found to be $-3.6$, quite
close to $-2\epsilon_{m}$ (-3.54). The Fr\"{o}hlich condition is
therefore nearly satisfied, so that the first-order surface mode
of the active nanoparticle is excited. As a consequence, the
nanoshell strongly scatters the incident light and significantly
localizes the electromagnetic field around its surface, as can be
found from the local field distribution (Fig. 2(c)). We also show
in Fig. 2(b) that a negative real permittivity $\epsilon_{b}=-4.2$
can be attributed to a composite with a filling fraction of 0.093.

To close the discussion, we should mention that $\epsilon_{b}$ can
be tuned by changing the filling fraction $f$, as suggested by Eq.
(\ref{eq5}). Taking the second design as an example, $f=0.15$
leads to $\epsilon_{b}=-3.9$, but $f=0.25$ gives
$\epsilon_{b}=-3.8$. In addition, it should be stressed that a
composite with compensated losses can easily be modified to
realize surface plasmon amplification by stimulated emission of
radiation (spaser)
\cite{Bergman,noginov2,zheludev,Noginov,Oulton,fang,stockman}. By
increasing the gain coefficient of the inclusive gain medium
slightly, it will begin to amplify light or even lase under proper
conditions. Therefore, the gain coefficient needed to compensate
for the loss of the localized surface plasmon can be set as a
threshold for the spaser \cite{noginov2}.

In conclusion, an electrostatic theory is developed to design bulk
composites with effective negative permittivities with compensated
losses. This theory is quite general and accounts for
nanoparticles with arbitrary shapes. Moreover, full-wave numerical
simulations were used to validate its accuracy. Two different
examples were considered to demonstrate the design procedure. It
was found that the first-order surface mode of the constitutive
nanoparticle should be excited to achieve a negative real-valued
$\epsilon_{b}$. The particle consequently has a zero absorption
cross section but a huge scattering cross section, strongly
enhancing its localized surface plasmon polariton and the
associated surface-enhanced Raman scattering.

\newpage
\begin{figure}[t]
\centering
\includegraphics[width=0.6\textwidth]{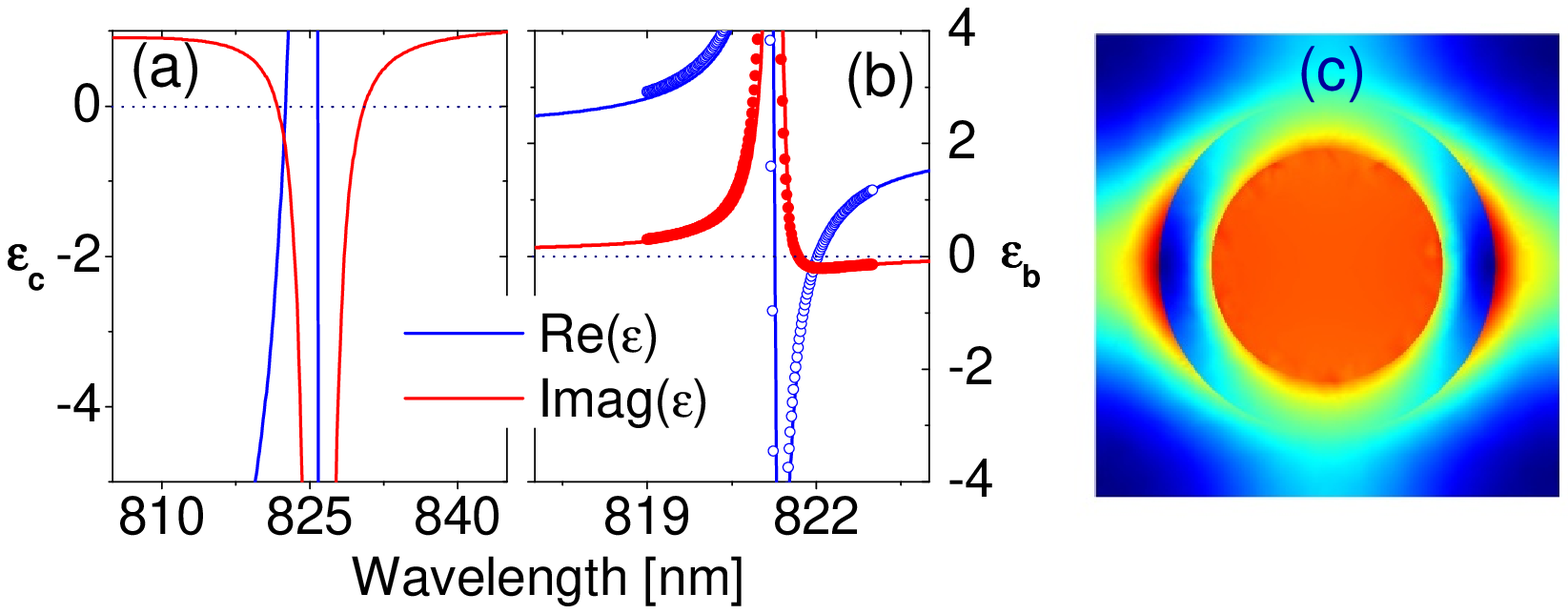}\vspace*{-8.5cm}
\caption{Spherical semiconductor quantum dots coated with a silver
film. (a) The effective permittivity $\epsilon_{c}$ of the
individual particle, and (b) the effective permittivity
$\epsilon_{b}$ of the bulk composite. The dotted curves are
obtained by full-wave simulations (see text). (c) The first-order
surface mode of the individual particle. The field magnitude is
color coded. The blue color corresponds to the minimum, while the
red color represents the maximum. The polarization of the incident
light is along the horizontal direction.} \label{fig1}
\end{figure}

\newpage
\begin{figure}[t]
\centering
\includegraphics[width=0.6\textwidth]{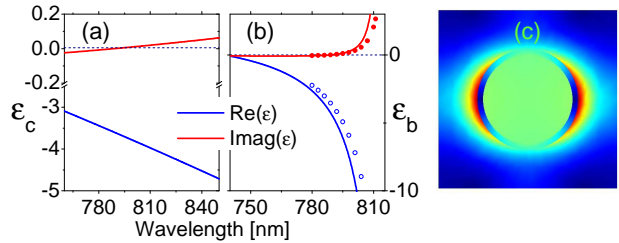}\vspace*{-8.5cm}
\caption{Same as Fig. 1, except the example is a gold coated gain
medium whose refractive index is $1.33-0.1437i$.} \label{fig2}
\end{figure}

\end{document}